\documentclass[superscriptaddress,twocolumn]{revtex4}
\usepackage{graphicx}
\usepackage{dcolumn}
\usepackage{bm}
\usepackage{amsmath}
\usepackage{amsfonts}

\begin{document}

\title{Born-Infeld Determinantal gravity and the taming of the conical singularity in 3-dimensional spacetime.}
\author{Rafael Ferraro}
\email{ferraro@iafe.uba.ar}
\thanks{Member of Carrera del Investigador Cient\'{\i}fico (CONICET,
Argentina)} \affiliation{Instituto de Astronom\'\i a y F\'\i sica
del Espacio, Casilla de Correo 67, Sucursal 28, 1428 Buenos Aires,
Argentina} \affiliation{Departamento de F\'\i sica, Facultad de
Ciencias Exactas y Naturales, Universidad de Buenos Aires, Ciudad
Universitaria, Pabell\'on I, 1428 Buenos Aires, Argentina}
\author{Franco Fiorini}
\email{franco@iafe.uba.ar} \affiliation{Instituto de Astronom\'\i
a y F\'\i sica del Espacio, Casilla de Correo 67, Sucursal 28,
1428 Buenos Aires, Argentina}

\begin{abstract}
In the context of Born-Infeld \emph{determinantal} gravity
formulated in a n-dimensional spacetime with absolute parallelism,
we found an exact 3-dimensional \emph{vacuum} circular symmetric
solution without cosmological constant consisting in a rotating
spacetime with non singular behavior. The space behaves at
infinity as the conical geometry typical of 3-dimensional General
Relativity without cosmological constant. However, the solution
has no conical singularity because the space ends at a minimal
circle that no freely falling particle can ever reach in a finite
proper time. The space is curved, but no divergences happen since
the curvature invariants vanish at both asymptotic limits.
Remarkably, this very mechanism also forbids the existence of
closed timelike curves in such a spacetime.
\end{abstract}


\maketitle

\section{Introduction}

Nowadays it is widely accepted by high energy physicists that
Einstein's theory must represent a low energy limit of a more
fundamental (quantum) theory of gravity. This suggests that the
transition between both regimes must be ruled by an ultraviolet
deformation of GR which, presumably, could solve many of the
puzzles present in Einstein's theory. In this direction, special
interest has been put on 3-dimensional gravity as an attempt to
understand many of the conceptual and technical problems
associated with the quantization of spacetime in the realistic
4-dimensional scenario \cite{Carlip}. In this process, it was
suddenly realized that three dimensional Einstein gravity has a
number of peculiar facts; it contains no propagating degrees of
freedom, and does not reduce to 2-dimensional Newtonian gravity in
the weak-field limit. Moreover, the spacetime is flat outside
matter and hence there exists no static interaction between
sources \cite{Brown}.

Not long after the first investigations in 3-dimensional General
Relativity have appeared \cite{Staru}, several generalizations
were proposed in order to make 3-dimensional dynamics more alike
the realistic $(3+1)$-gravity. Among the plethora of theories that
are not constrained to exist only in 3 dimensions we can mention
$(2+1)$-dilaton gravity \cite{Dilaton}-\cite{Dilaton2}, conformal
gravity \cite{Conformal1,Conformal2} and the newcomer New Massive
Gravity (NMG) \cite{Town1,Town2}. On the other hand, some
constructions that are unique to $2+1$ dimensions have been also
considered. One that has attracted much attention in the last
years is the so called Topological Massive Gravity (TMG), which
adds to Einstein action a Chern-Simons term free of torsion
\cite{TMG1,TMG2} (see \cite{TMG4} for a comprehensive review of
solutions).

The singularities inherent to Einstein theory had been matter of
research since the early days of General Relativity (GR). Thought
the concept of singularity encounter its \emph{raison d'\^{e}tre}
in the geodesic incompleteness \cite{H-E}, it historically came
into light associated with the divergences of physical quantities.
Regarding this matter, most of the major achievements in the
subject have arisen from examination of two fundamental issues:
the question of the origin of the Universe and the final state
occurring in the gravitational collapse of massive stars. In the
former issue (leaving aside ontological discussions about the
origin of time), physical quantities such as the energy density
and pressure of matter fields, become infinite in the Big Bang. In
the latter, the unfortunate destiny of the infalling observer who
goes beyond the Schwarzschild radius, is to experiment infinite
tidal forces as he/she approaches $r=0$, due to the very
infiniteness of the Riemann curvature tensor at that point.

As is well known, vacuum solutions for 3-dimensional GR are free
of curvature singularities, because the Einstein tensor is just
the double dual of the curvature and so, essentially, it is
proportional to the stress-energy tensor. However, due to non
trivial topological properties, the massive circular symmetric
solutions of vacuum Einstein equations in 3 spacetime dimensions
displays a conical singularity at the origin. In the case without
cosmological constant, which might be considered the three
dimensional analogue of the exterior Kerr metric, the solution
exhibits closed timelike curves (CTC). More realistic four
dimensional cosmic string solutions inherit all these properties
\cite{Vil,Kib}.

In papers \cite{Nos,Nos2} we have introduced the so called
Born-Infeld (BI) gravity with the aim of smoothing the curvature
singularities characterizing the cosmological
(Friedmann-Robertson-Walker) solutions of $n=3+1$ GR. In the
present work, in turn, we are pursuing a different task by asking
whether it is possible to remove the singularities of topological
nature existent in vacuum $(2+1)$ Einstein Gravity. It is worth of
mention that none of the above referred approaches to gravity in 3
dimensions have supplied a non-singular behavior in its circular
symmetric vacuum solutions. For this purpose we extend the
construction presented in the articles \cite{Nos,Nos2} by working
with a determinantal form of the action. This new approach to the
subject has the benefit of being in more close correspondence with
the original BI construction. For this new scheme we have found a
circular symmetric vacuum solution in three dimensional spacetime
without cosmological constant. We have obtained that the angular
momentum $J$ not only controls the global properties of the
spacetime, but it has an impact on the local physics through the
curvature of the manifold. Remarkably, the curvature invariants
are bounded functions of the radial coordinate. When the BI
parameter $\lambda$ tends to infinity, the conical geometry
characterizing the elementary solution of Einstein's theory in
$n=3$ is restored. Particularly interesting is the fact that the
theory provides a minimum attainable circle whose circumference is
$\pi J/M$, where $M$ is a constant related with the mass of the
spinning source. As a consequence, the spacetime structure becomes
geodesically complete because no free falling particles can ever
reach this minimum circle in a finite proper time. Another feature
of this natural cutoff on the radial coordinate is that, unlike
its low energy (i.e. GR) version, there are not closed timelike
curves in this geometry.

\section{Born-Infeld gravity in Weitzenb\"{o}ck spacetime}

In order to motivate the construction we will work out, let us
briefly examine Born-Infeld electrodynamics. As is well known,
this non linear theory for the electromagnetic field was able to
tame the infinite self energy of the point-like charged particle.
In its first version \cite{Born,Borna}, BI theory deformed the
Maxwell Lagrangian $L_{M}\propto( \mathbf{E}^2- \mathbf{B}^2)$
according to the rule
\begin{equation}  \label{scheme}
\mathcal{I}_{M}\longrightarrow\mathcal{I}_{BI0}=\lambda\int d^4x\,
\left[\sqrt{1+\lambda^{-1}L_{M}}-1\right].
\end{equation}
The scheme (\ref{scheme}) is not as unnatural as it seems at first
glance; the same technique can be used for going from the
classical free particle action to the relativistic one; in such
case, the scale is $\lambda=-m c^2$, which smoothes the particle
velocity by preventing its unlimited growing. In the regime where
$L=m\mathbf{V}^2<<\lambda$ the relativistic physics restore its
low energy (Newtonian) realm.

Soon after its advent \cite{BornI}-\cite{BornIII}, Born and Infeld
generalized their construction by considering the generally
covariant determinantal action
\begin{equation}
\mathcal{I}_{\mathbf{BI}}=\lambda \int d^{4}x\Big[\sqrt{|g_{\mu
\nu }+\lambda ^{-1}F_{\mu \nu }|}-\sqrt{|g_{\mu \nu }|}\Big],
\label{acciondetelectro}
\end{equation}
which implicitly includes also the pseudo-invariant
$\mathbf{E}\cdot \mathbf{B }$ ($|\ \ |$ stands for the absolute
value of the determinant). Expressions (\ref{scheme}) and
(\ref{acciondetelectro}) are coincident only in pure electrostatic
or magnetostatic situations, or in electrodynamical phenomena
concerning plane waves (where the two field invariants are null).
In this last case, the scale $\lambda $ plays not role at all,
hence the field configurations are exactly the same than those of
Maxwell's theory. BI electrodynamics reduces to Maxwell's theory
for small amplitudes, both of them having causal propagation and
absence of birefringence. Remarkably, after a long exile, BI
action came back again to the stage in the context of more modern
developments; the quartic terms implicit in
(\ref{acciondetelectro}) reproduce the effective action of
one-loop supersymmetric QED \cite{Tsey1}, and the structure
(\ref{acciondetelectro}) emerge naturally in the low energy limit
of string theory as the action governing the electromagnetic field
of D-branes \cite{Tsey2}.

The above mentioned remarkable features of the BI program,
together with its well known curative properties concerning
singularities, invites to search for gravitational analogues with
the structure (\ref{acciondetelectro}). This matter has attracted
some attention in the past \cite{deser3}-\cite{Wohlfarth}, where
several deformations \`{a} la Born-Infeld combining higher order
invariants related to the curvature in a Riemannian context were
tried. More recently, a thorough analysis of cosmological models
by means of dynamical systems techniques was performed in
\cite{Quiros}. All these constructions, however, lead to
troublesome four order field equations for the metric. Actually,
within these frameworks, exact solutions were never found. In
spite of this, the importance of BI-like actions for the
gravitational field was revisited very recently in connection with
the problem of quantum gravity \cite{Tek1,Tek2}. In a different
direction, BI-like actions were explored also in Refs.
\cite{Vollick}, \cite{Max} and \cite{Max1} using the Palatini
formalism, where metric and connection are taken as independent
entities. In this article, we shall follow a different path by
considering a BI deformation in Weitzenb\"{o}ck spacetime.

General Relativity can be formulated in a spacetime possessing
absolute parallelism. This approach is usually known as
teleparallel equivalent of General Relativity TEGR
\cite{Hehl,Hehl2}, and relies on the existence of a set
$\{e^{a}(x)\}$ of $n$ one-forms that turn out to be autoparallel
for the Weitzenb\"{o}ck connection $\Gamma _{\mu \nu }^{\lambda
}=e_{a}^{\lambda }\,\partial _{\nu }e_{\mu }^{a}$ ($e_{a}^{\lambda
}$ makes up the inverse matrix of $e_{\mu }^{a}$). This connection
is compatible with the metric $\mathrm{g}(x)=\eta _{ab}\
e^{a}(x)\otimes e^{b}(x)$ and curvature free: Weitzenb\"{o}ck
spacetime is flat though it possesses torsion $T^{a}=de^{a}$,
which is the agent where the gravitational degrees of freedom are
encoded. The structure of the torsion tensor resembles the one of
the electromagnetic field tensor $F=dA$ and, like Maxwell's,
teleparallel Lagrangian density is quadratic in this tensor. In
fact, TEGR action with cosmological constant $\Lambda $ is
\cite{Maluf}
\begin{equation}
\mathcal{I}_{\mathbf{GR||}}=\frac{1}{16\pi G}\int
d^{n}x\,\sqrt{|g_{\mu \nu }|}\,(\mathbb{S}\cdot
\mathbb{T}-2\Lambda ),  \label{acciongr}
\end{equation}%
where $\mathbb{S}\cdot \mathbb{T}\doteq S_{\rho }^{\ \ \mu \nu
}\,T_{\ \ \mu \nu }^{\rho }$, $T_{\ \ \mu \nu }^{\rho
}=e_{a}^{\rho }\,(\partial _{\mu }e_{\nu }^{a}-\partial _{\nu
}e_{\mu }^{a})$ and $S_{\rho }^{\ \ \mu \nu }$ is defined as
\begin{equation}
S_{\rho }^{\ \ \mu \nu }=-\frac{1}{4}\,(T_{\ \ \ \rho }^{\mu \nu
}-T_{\ \ \ \rho }^{\nu \mu }-T_{\rho }^{\ \ \mu \nu
})+\frac{1}{2}(\delta _{\rho }^{\mu }\,T_{\ \ \ \theta }^{\theta
\nu }-\delta _{\rho }^{\nu }\,T_{\ \ \ \theta }^{\theta \mu }).
\notag  \label{tensorS}
\end{equation}%
The equivalence between GR and the theory (\ref{acciongr}) comes
from the fact that the GR Lagrangian -i.e. the curvature scalar
$R$ of the Levi-Civita connection- is $R=\mathbb{S}\cdot
\mathbb{T}+$ \emph{Surface Terms}. In this expression, the surface
terms encompass all the second derivatives entering the scalar
curvature $R$. In fact, Weitzenb\"{o}ck torsion $T$ contains just
first derivatives of the fields $e^{a}(x)$. This distinctive
feature makes Weitzenb\"{o}ck torsion a privileged geometric
structure to formulate modified theories of gravitation, since it
guarantees that any modified Lagrangian in this language will
assure second order field equations.

In Ref. \cite{Nos} we followed the spirit of Eq.~(\ref{scheme}) by
studying the \emph{deformed} action
\begin{equation}
\mathcal{I}_{\mathbf{BI0}}=\lambda \int d^{n}x\,\sqrt{|g_{\mu \nu }|}\Big[%
\sqrt{1+2\lambda ^{-1}\ \mathbb{S}\cdot \mathbb{T}}-\alpha \Big],
\label{accionBI0}
\end{equation}
which proved to be capable of smoothing the GR cosmological
singularity, providing a natural inflationary stage (without the
mediation of an inflaton) and bounding the dynamics of the Hubble
parameter \footnote{See Ref.~\cite{Nos3} for a brief summary of
these results in 4 dimensions}. Apart from this cosmological
success, the scheme (\ref{accionBI0}) was unable to deform the
3-dimensional circular symmetric solutions, in particular the BTZ
black hole \cite{btz}. This inability is a consequence of the fact
that the scalar Lagrangian in (\ref{accionBI0}) is constant on the
BTZ solution: $\mathbb{S}\cdot \mathbb{T}=-2\Lambda $ \cite{Nos2}.

Here we will follow the spirit of (\ref{acciondetelectro}), so we
shall propose the general n-dimensional BI action in
Weitzenb\"{o}ck spacetime
\begin{equation}
\mathcal{I}_{\mathbf{BIG}}=\frac{\lambda /(A\!+\!B)}{16\pi G}\!\int \!d^{n}x%
\left[ \sqrt{|g_{\mu \nu }+2\lambda ^{-1}\mathcal{F}_{\mu \nu
}|}-\alpha \sqrt{|g_{\mu \nu }|}\right],   \label{acciondet}
\end{equation}%
where $\mathcal{F}_{\mu \nu }$ is quadratic in the Weitzenb\"{o}ck
torsion, and reads $\mathcal{F}_{\mu \nu }=A\,S_{\mu \lambda \rho
}T_{\nu }^{\,\,\,\lambda \rho }+B\,S_{\lambda \mu \rho
}T_{\,\,\,\,\,\nu }^{\lambda \,\,\,\,\rho }$, $A$ and $B$ being
non-dimensional constants. Such a
combination ensures the correct GR limit since both terms in $\mathcal{F}%
_{\mu \nu }$ have trace proportional to $\mathbb{S}\cdot
\mathbb{T}$. In
fact, we can factor out $\sqrt{|g_{\mu \nu }|}$ from expression (\ref%
{acciondet}) and use the expansion of the determinant,
\begin{equation}
\det (\mathbb{I}-\epsilon \,\mathbb{F})=1+p_{1}\,\epsilon
+p_{2}\,\epsilon ^{2}+...+p_{n-1}\,\epsilon ^{n-1}+p_{n}\,\epsilon
^{n},\notag
\end{equation}%
where
\begin{eqnarray}
p_{1} &=&-s_{1}  \notag \\
p_{2} &=&-\frac{1}{2}(s_{2}+p_{1}s_{1})  \notag \\
. &&.  \notag \\
p_{n} &=&-\frac{1}{n}(s_{n}+p_{1}s_{n-1}+...+p_{n-1}s_{1}),
\notag
\end{eqnarray}%
and $s_{i}=Tr(\mathbb{F}^{i})$. In our case it is $\epsilon =-2\lambda ^{-1}$%
\ and $\mathbb{F}\equiv \mathcal{F}_{\mu }^{\,\,\nu }$. Thus the
Lagrangian density in $\mathcal{I}_{\mathbf{BIG}}$ is

\begin{widetext}

\begin{eqnarray*}
\mathcal{L}_{\mathbf{BIG}} &=&\frac{\lambda /(A\!+\!B)}{16\pi
G}\sqrt{|g_{\mu \nu }|}\,\Big[ 1+\lambda ^{-1}\mathcal{F}_{\mu
}^{\,\,\mu }
+\lambda ^{-2}\left( \frac{1}{2}(\mathcal{F}_{\mu }^{\,\,\mu })^{2}-\mathcal{%
F}_{\mu }^{\,\,\nu }\mathcal{F}_{\nu }^{\,\,\mu }\right) -\alpha \Big] +%
\mathcal{O}(\lambda ^{-2}) \\
&=&\frac{\sqrt{|g_{\mu \nu }|}}{16\pi G}\;\Big[ \mathbb{S}\cdot \mathbb{T+\,}%
\frac{A\!+\!B}{2\lambda }(\mathbb{S}\cdot \mathbb{T})^{2}
-\frac{1}{\lambda (A\!+\!B)}\mathcal{F}_{\mu }^{\,\,\nu
}\mathcal{F}_{\nu }^{\,\,\mu }-\frac{\lambda (\alpha
-1)}{A\!+\!B}\Big] +\mathcal{O}(\lambda ^{-2}).
\end{eqnarray*}%
\end{widetext}

At the lowest order we retrieve the low energy regime described by
the Einstein theory (\ref{acciongr}) with cosmological constant
$\Lambda
=\lambda (\alpha -1)/[2(A+B)]$. The following term $\lambda ^{-1}(\mathbb{S}%
\cdot \mathbb{T})^{2}$ is also present in the expansion of action (\ref%
{accionBI0}). However we get now a new term $\mathcal{F}_{\mu }^{\,\,\nu }%
\mathcal{F}_{\nu }^{\,\,\mu }$ at the order $\lambda ^{-1}$, so $\mathcal{I}%
_{\mathbf{BIG}}$ departs from $\mathcal{I}_{\mathbf{BI0}}$ even at
the order $\lambda ^{-1}$. Whether the action (\ref{acciondet})
can be regarded as an effective (low energy) action for gravity
coming from a more fundamental quantum theory is unknown at
present, perhaps because the very quantum theory of gravity is yet
a tale to be unfolded. Nevertheless, the experience
acquired with its electromagnetic analogue suggest that theory (\ref%
{acciondet}) would constitute a slope worth to be explored. Action (\ref%
{acciondet}) shows us that the framework (\ref{accionBI0}) is,
among the whole Born-Infeld catalogue, just the top of the
iceberg. The use of a
Lagrangian which is not a mere deformation of the one in action (\ref%
{acciongr}) opens the possibility of finding a high energy
modification for the GR spherically symmetric solutions. In the
next section we show that this is indeed the case.

\section{Taming the conical singularity and erasing CTC's.}
We will investigate the properties of action (\ref{acciondet}) in
the more accessible environment of (2+1)-gravity. In particular,
let us work under the assumption of spherically (circular)
symmetric spacetimes, and propose the following driebein written
down in standard polar coordinates $(t,r,\theta )$
\begin{eqnarray}
e^{0} &=&N(r)dt,  \notag \\ e^{1} &=&(Y(r)/N(r))dr, \label{triada}
\\ e^{2} &=&r(N^{\theta }(r)\,dt+\,d\theta ), \notag
\end{eqnarray}%
which implies the metric tensor
\begin{equation}
ds^{2}=N^{2}(r)\
dt^{2}-\frac{Y^{2}(r)}{N^{2}(r)}dr^{2}-r^{2}\left( N^{\theta }(r)\
dt+d\theta \right) ^{2}.  \notag  \label{metrica}
\end{equation}%
As is known, the vacuum solution for the GR ($\lambda \rightarrow
\infty $)
limit is%
\begin{equation}
N_{o}^{\theta }(r)=-\frac{J}{2\,r^{2}}\;,\ \ \
N_{o}^{2}(r)=-M-\Lambda \,r^{2}+\frac{J^{2}}{4\,r^{2}}\;,\,\ \
Y=1\   \label{undefor}
\end{equation}%
which becomes the rotating BTZ black hole when $\Lambda <0$.

We will try the dreiben (\ref{triada}) in the dynamical equations
coming from the action (\ref{acciondet}), for the particular case
$B=0$ (constant $A$ will be absorbed in $\lambda$). In terms of
the natural variables defined as
\begin{equation}
X=-\frac{(N^{2})^{\prime }}{\lambda \,r\,Y^{2}}\;,\ \ \ \ \ \ \ \,\,\,\,Z=%
\frac{r^{2}(N^{\theta \,\,^{\prime }})^{2}}{2\lambda \,Y^{2}},
\label{varnat}
\end{equation}%
the dynamical equations read
\begin{eqnarray}
\frac{1-X+Z/2}{\sqrt{\mathcal{U}(X,Z)}} &=&K\,Y,  \label{equis1}
\\ \frac{\sqrt{2\lambda \,Z}(1-X/2)}{\sqrt{\mathcal{U}(X,Z)}}
&=&\frac{J}{r^{2}} ,  \label{y1} \\ (1+2\,\Lambda /\lambda
)\,\sqrt{\mathcal{U}(X,Z)} &=&1-X^{2}+XZ, \label{zeta1}
\end{eqnarray}%
with
\begin{equation}
\mathcal{U}(X,Z)=1-2X+X^{2}+2Z-ZX,  \label{U}
\end{equation}%
$K$ and $J$ being two integration constants. Actually $K$ can be
absorbed in $Y$ by redefining the variables $N$, $N^{\theta }$ and
the coordinate $t$ (without affecting $X$, $Z$); so, we will use
$K=1$. Equations (\ref{equis1})-(\ref{zeta1}) are three coupled
algebraic equations. In spite of its apparent harmlessness, they
are quite hard to solve in its full generality.

In the case $\Lambda =0$ it is not difficult to find an exact
solution for
the system (\ref{equis1})-(\ref{zeta1}). Notice that the GR solution (\ref%
{undefor}) satisfies the relation $X=Z$, which in turn leads to $\mathcal{U}%
(X=Z,Z)=1$ (see Eq.~(\ref{U})). If $\Lambda =0$ then $\lambda $
does not explicitly appear in Eq.~(\ref{zeta1}). So the relation
$X=Z$ is still suitable to solve Eq.~(\ref{zeta1}). The remaining
equations are cast in the form
\begin{eqnarray}
1-\frac{Z}{2} &=&Y,  \label{X=-Z} \\
Z\,Y^{2} &=&\frac{J^{2}}{2\lambda \,r^{4}}\doteq 2\Delta.
\label{X=-Z1}
\end{eqnarray}%
From  Eq.~(\ref{X=-Z1}), the definitions (\ref{varnat}) for $X$,
$Z$ and the relation $X=Z$ one gets:
\begin{equation}
N^{\theta }(r)=-\frac{J}{2\,r^{2}}\;,\ \ \ \ \ \ \ \ N^{2}(r)=M^{2}+\frac{%
J^{2}}{4\,r^{2}},  \label{resulh}
\end{equation}%
where $M^{2}$ is an integration constant. Thus the interval takes the form%
\begin{widetext}
\begin{eqnarray}
ds^{2}&=&\left( J^{2}/(4\,r^{2})+M^{2}\right) \ dt^{2}
-\left( \frac{Y(r)^{2}}{J^{2}/(4\,r^{2})+M^{2}}\right) dr^{2}-r^{2}\left( -%
\frac{J}{2\,r^{2}}\ dt+d\theta \right) ^{2}  \notag \\
&=&\left[ d(M\ t+J\ \theta /(2M))\right] ^{2}-\left( \frac{Y(r)^{2}}{%
J^{2}/(4\,r^{2})+M^{2}}\right) dr^{2}
-\frac{r^{2}}{M^{2}}(J^{2}/(4r^{2})+M^{2})\ d\theta ^{2}.
\label{metricareg}
\end{eqnarray}%
\end{widetext} By performing the changes
\begin{eqnarray}
r &\rightarrow
&\rho=M^{-2}(J^{2}/4+M^{2}r^{2})^{1/2},\label{cambios1}
\\
t &\rightarrow &T=M\ t+J\ \theta /(2M),\label{cambios2}
\end{eqnarray}%
the interval (\ref{metricareg}) is cast in the form
\begin{equation}
ds^{2}=dT^{2}-Y(\rho)^{2}\ d\rho^{2}-M^{2}\rho^{2}d\theta ^{2}.
\label{metcono}
\end{equation}
In the TEGR limit ($\lambda \rightarrow \infty $) it is $\Delta
\rightarrow 0$; then $Z\rightarrow 0$ and $Y\rightarrow 1$ in
Eqs.~(\ref{X=-Z})-(\ref{X=-Z1}). Thus the flat spacetime is
locally recovered (notice that the constant $M$ could be absorbed
by redefining $\theta$). From a global viewpoint,
Eq.~(\ref{metcono}) with $Y=1$ could be regarded as a conical
structure: the slices $T=constant$, $0\le\theta<2\pi$,
$0\le\rho<\infty$ are planes where a wedge was cut off and its
opposite sides were identified. The deficit angle is $\beta =2\pi
(1-M)$ ($\beta$ is related with the mass $m$ of a source at the
origin: $m=\beta/(2\pi G)$ \cite{3D}). Actually, the coordinate
$\rho$ is not allowed to reach the value $\rho=0$ in
Eq.~(\ref{cambios1}). However, in TEGR this is not a real
limitation of coordinate $\rho$ but a consequence of the chosen
dreibein. In fact $\rho$ can be effectively extended up to
$\rho=0$, as is apparent in Eq. (\ref{metcono}) with $Y=1$.

As is well known, TEGR theory (\ref{acciongr}) is invariant under
local Lorentz transformations of the vielbein; therefore the
geometry (\ref{metcono}) with $Y=1$ could be derived not only from
the dreibein (\ref{triada}) but from the inertial dreibein
$\{E^0=dT,\ E^1=d\rho=e^1, E^2=M\rho\, d\theta \}$. Both dreibeins
are related by the Lorentz transformation
\begin{equation}
e^0=\frac{E^0-J/(2M^2\rho)\,E^2}{\sqrt{1-J^2/(4M^4\,\rho^2)}},
\,\,\,\,\,\,\,
e^2=\frac{E^2-J/(2M^2\rho)\,E^0}{\sqrt{1-J^2/(4M^4\,\rho^2)}}\notag,
\end{equation}
which is a boost tangent to the circle $\rho=constant$ with
velocity $V=J/(2M^2\,\rho)$. So $J$ measures the rotation of the
dreibein (\ref{triada}) with respect to the inertial frame. The
boost velocity increases from infinity to reach the maximum value
at $\rho=J/(2M^2)$, i.e. at $r=0$ (see Eq.~(\ref{cambios1})).
However, as a consequence of the gauge freedom, the geometry
(\ref{metcono}) with $Y=1$ is not imprinted with the value of $J$.
Thus one can fix the gauge by choosing $J=0$, which amounts to the
choice of the inertial dreibein, so extending the range of $\rho$
from infinity up to zero.

On the contrary, the modified teleparallel actions are invariant
only under \textit{global} Lorentz transformations of the vielbein
\cite{Nos}, which pre-announce a different role of $J$ in these
theories and a geometrical meaning for the bound $\rho> J/(2M^2)$.
In fact, whereas a local Lorentz transformation of the vielbein
adds a divergence term to $\mathbb{S}\cdot \mathbb{T}$, which is
not physically significant in action (\ref{acciongr}), instead
such a divergence term does affect the modified actions
(\ref{accionBI0}) and (\ref{acciondet}). This loss of gauge
freedom means that the modified teleparallel theories govern more
dynamical variables that TEGR does. Thus, the parameters
characterizing the lost gauge transformations become integration
constants associated with the recovered degrees of freedom.
Therefore, the family of metrics resulting from the solutions is
enlarged. Because of this, $J$ plays a very different role in
modified teleparallelism; since dreibeins with different values of
$J$ are not related through global Lorentz transformations, then
they represent genuine different solutions of the theory. $J$
enters the metric (\ref{metcono}) to make $Y$ a function of the
radial coordinate, so labeling different (curved) solutions. In
fact, according to Eqs.~(\ref{X=-Z})-(\ref{X=-Z1}), the function
$Y(r)$ is obtained from the cubic equation
\begin{equation}
Y^{2}-Y^{3}=\frac{J^2}{4\,\lambda\, r^4}=\Delta.\label{cubic}
\end{equation}
In the modified theory (i.e., for finite values of $\lambda$), the
flat solution $Y=1$ can only be obtained when $J=0$, otherwise the
space is curved. The integration constant $J$ is the source of the
curvature. According to Eq.~(\ref{cubic}), $J/\sqrt{|\lambda |}$
is the squared length scale for such deformation of flat
spacetime. Alternatively, the spatial curvature could be regarded
as a variable deficit angle (just perform the coordinate change
$d\xi=Y(\rho)\, d\rho$ in (\ref{metcono})). Summarizing, the
modified theory not only contains the GR solution but a family of
curved spacetimes parametrized by the integration constant $J$. As
we are going to show, the curvature of the solutions with $J\neq
0$ softens the conical singularity by replacing it with an
unreachable minimal circle of radius $\rho_o\equiv J/(2M^2)$.

Among the three solutions of Eq.~(\ref{cubic}), we will keep the
one going to $1$ when $\Delta \rightarrow 0$, since it contains
both the GR limit and the proper behavior at infinity. This
solution is:
\begin{eqnarray}\label{zetarara}
3\ Y&=&1+\left( 1-\frac{27}{2}\ \Delta -\frac{3}{2}\sqrt{3\Delta
\,(27\Delta -4)}\right) ^{-1/3}+\notag \\
&+&\left(1-\frac{27}{2}\ \Delta -\frac{3}{2}\sqrt{3\Delta
\,(27\Delta -4)} \right) ^{1/3}.
\end{eqnarray}
If $\lambda <0$ then $\Delta<0$ and the function $Y$ is defined
for $0<r<\infty$ (i.e., $J/(2M^{2})<\rho<\infty$); so hereafter we
shall focus in the case with $\lambda<0$. We can characterize the
geometry (\ref{metcono}) by computing its curvature invariants:
\begin{eqnarray}
R&=&\frac{2Y(\rho)'}{\rho Y(\rho)^3}=\frac{2Y(r)'}{r Y(r)^3},
\,\,\,\,\,R^{\mu\nu}R_{\mu\nu}=\frac{1}{2}\, R^2, \notag\\
K&\equiv& R^\alpha_{\,\,\,\beta\gamma\delta}\,
R_{\alpha}^{\,\,\,\beta\gamma\delta}=R^2 \label{inva1}.
\end{eqnarray}
Of course, they go to zero for $r$ (or $\rho$) going to infinity.
In this case, due to the fact that $Y\rightarrow1$ when
$\rho\rightarrow\infty$, the metric (\ref{metcono}) describes the
conical (locally flat) GR spacetime.

\begin{figure}[ht]
\centering \includegraphics[scale=.65]{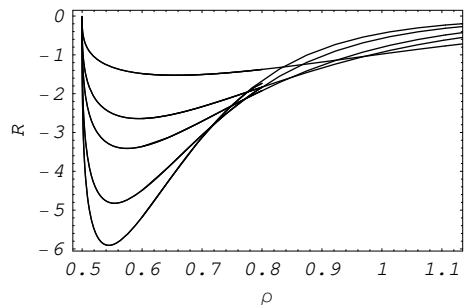}
\caption{Scalar curvature $R$ as a function of the radial
coordinate $\rho$, for $J/M=1$. Following the minimum of the
curves from bottom to top, it is $-\lambda=15,10,5,3,1$.}
\label{Curvas}
\end{figure}

The invariants (\ref{inva1}) also go to zero for $r\rightarrow 0$
(or $\rho\rightarrow J/(2 M^2)$). In fact, according to
Eq.~(\ref{cubic}) $Y$ behaves as $(-\Delta)^{1/3}$ for
$\Delta\rightarrow\infty$, what implies
\begin{equation}
R \sim -\frac{16}{3}\, \left(\frac{\sqrt{2}\, |\lambda|\,
r}{J^2}\right)^{2/3},
\end{equation}
when $r\rightarrow 0$. As was said before, the coordinate change
$d\xi=Y(\rho)\, d\rho$ in the metric (\ref{metcono}) allows to
regard this curved geometry as a space of a variable deficit angle
ranging from $2\pi(1-M)$ at spatial infinity, to $2\pi$ at
$\rho_o$ ($r=0$). In this last limit, since the deficit cover the
whole range of the angular variable, the metric describes a
cylinder of radius $\rho_{o}$, which is obtained by identifying
points in opposite sides of the \emph{total} wedge. Thus, the
geometry (\ref{metcono}) is also asymptotically locally flat when
$\rho\rightarrow\rho_{o}$. Figure (\ref{Curvas}) depicts the
scalar curvature $R(\rho)$, for $J/M=1$ and several negative
values of the Born-Infeld parameter $\lambda$. The minimum
curvature is reached at a position $\rho_{min}$ that depends only
on the combination $\mathtt{J}=J/ \sqrt{|\lambda|}$. In the highly
deformed regime $\mathtt{J}\approx 1$ the curvature effects can be
felt at positions very distant from the origin. Instead, as long
as the low energy limit is restored ($\mathtt{J}<<1$), such
effects are confined to small neighborhoods of $\rho_{o}=J/2M^2$
(i.e. $r=0$). In the GR limit $\lambda\rightarrow\infty$
(equivalently, $Y\rightarrow1$) there are no effect at all,
because the manifold becomes flat.

\begin{figure}[ht]
\centering \includegraphics[scale=.35]{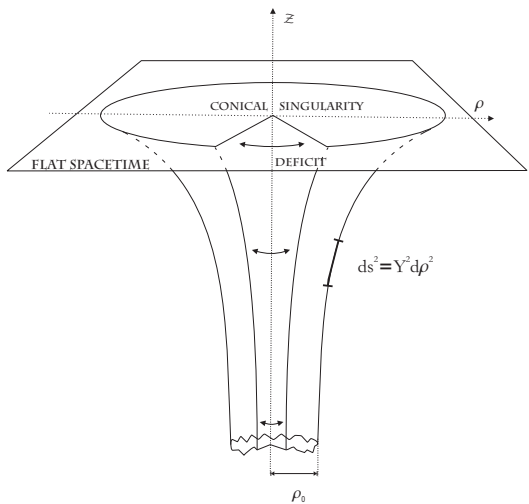}
\caption{Schematic representation of the spacetime (\ref{metcono})
as embedded in three-dimensional Euclidean space.}
\label{geomconi}
\end{figure}

The lower bound for the radial coordinate means that the space
ends at a minimal circle of circumference $2\pi\, M\, \rho_{o}=
\pi\, J/M$. However this boundary requires an infinite proper time
to be reached, which implies that the conical singularity is
smoothed. In fact the radial light rays satisfy $dT=Y d\rho$, so
the coordinate time $T$ diverges when a light ray approaches the
minimal circle (because $Y$ diverges). On the other hand, since
the metric components in Eq.~(\ref{metcono}) do not depend on $T$,
then $p_T = g_{TT}\, p^T = p^T \propto dT/d\tau$ is conserved on
geodesics. This means that the proper time $\tau$ of a freely
falling particle is proportional to the coordinate time $T$. Since
timelike geodesics remain inside the light cones, then a particle
needs an infinite proper time to reach the minimal circle. In
Figure (\ref{geomconi}) we have schematically depicted the
spacetime (\ref{metcono}) with $T=constant$ as embedded in three
dimensional Euclidean space with coordinates $(\rho,\theta,z)$.
The \emph{funnel}-like structure appearing in the figure comes
from the function $z(\rho)$ which is
\begin{equation}
z=\int\sqrt{Y^2(\rho)-1}\ d\rho,
\end{equation}
so the Euclidean squared interval on that curve is given by
$ds^2=dz^2+d\rho^2=Y^2d\rho^2$. In the asymptotic region we have
$Y\rightarrow 1$ and then, $z$ becomes constant there (we set
$z=0$ in the figure).

The obtained geometry not only succeeds in smoothing the conical
singularity of the GR ($Y=1$) solution but avoids another
unpleasant feature of Einstein theory in $n=2+1$ that was posed in
early works \cite{3D}-\cite{3D3}: the existence of closed timelike
curves (see also Ref.~\cite{3D4} where additional physical
criteria was discussed in order to avoid CTC). Such a undesirable
property appears when coordinate $t$ is considered continuous
instead of $T$. This condition forces a jump $\Delta T=J\pi/M$
along the circle ($\Delta\theta=2\pi$). While a jump of $\theta$
(deficit angle) is related with the mass $m$ of the solution, a
jump of $T$ provides the solution with angular momentum. In fact,
by replacing the solution (\ref{metcono}) whit $Y=1$ in
(2+1)-Einstein equations it results that the energy-momentum
tensor of the source is $T^{tt}\propto m\, \delta^2({\bf r})$,
$T^{ti}\propto J\, \varepsilon^{ij}\,
\partial_j \delta^2({\bf r})$; i.e., a spinning massive particle
is at the origin $\rho=0$ \cite{3D}. In this spacetime we can
consider the closed curve with constant $(t,\rho)$ in the interval
(\ref{metricareg}) under the coordinate change given in
(\ref{cambios1}). It then becomes
\begin{equation}
ds^{2}=\left[\left(\frac{J}{2 M^2}\right)^2-\rho^2\right]\, M^2\,
d\theta^2.
\end{equation}
For $\rho<J/(2M^2)$ the closed curve in $\theta$ would be
time-like. GR allows this possibility, since $Y=1$ and no
restrictions appears for the coordinate $\rho$. In the
determinantal theory, instead, $\rho$ is constrained to be greater
than $J/(2M^2)$, so excluding CTC. The same mechanism responsible
for the taming of the conical singularity at the origin seems to
provide a natural chronological protection.

\section{Concluding comments}
Born-Infeld determinantal action (\ref{acciondet}) could be seen
as a natural ultraviolet deformation of Einstein gravity which
operates at scales of order $\ell\sim|\lambda|^{-1/2}$. For the
theory (\ref{accionBI0}), which could be considered the simplest
structure among the BI program, it was shown in Ref.~\cite{Nos2}
that this scale plays an important role in $n$-dimensional
cosmological scenarios, because it works as an effective initial
vacuum energy driving the inflationary stage. Moreover, the
invariants are bounded by the BI parameter $\lambda$, ruling in
this way not only the behavior of the inflationary phase, but also
establishing a maximum attainable spacetime curvature, with its
subsequent singularity avoidance.

In the present context we witness a similar behavior; while action
(\ref{accionBI0}) was unable of deforming three dimensional vacuum
solutions, its extension (\ref{acciondet}) contains non constant
curvature states in empty space. The example considered here, the
one given by metric (\ref{metricareg}), is particularly
interesting because it represents a circular symmetric spacetime
with bounded curvature invariants and $\alpha=1$, i.e., without
cosmological constant. The relevant parameter in the deformation
is $\mathtt{J}=J/\sqrt{|\lambda|}$, so extremely high energy
regimes leads to strongly rotating systems
($J^2=\mathcal{O}(|\lambda|)$). The asymptotic spacetime is the
conical geometry (\ref{metcono}) with $Y=1$ typical of three
dimensional GR solutions without cosmological constant. However,
while the GR solution has $Y=1$ $\forall \rho$, the determinantal
action leads to the behavior (\ref{zetarara}) for the function
$Y$. In this way the singularity is removed and replaced with an
unreachable asymptotic minimal circle. Both asymptotic regions are
flat, but the space between them is curved. So, unlike GR, the
angular momentum not only affects the global properties of the
spacetime, but also has an effect on its curvature. Furthermore,
its presence is crucial in order to erase the conical singularity
at the origin, and to give rise an spacetime free of CTC.

It is worth mentioning that the results here obtained are clearly
extensible to the four dimensional cosmic string solution, whose
metric reads
\begin{equation}
ds^{2}=dT^{2}-Y(\rho)^{2}\ d\rho^{2}-M^{2}\rho^{2}d\theta
^{2}-dz^2, \label{metcuerda}
\end{equation}
where now the slices $T=constant$ are described in cylindrical
coordinates $(\rho,\theta,z)$. As another remarkable physical
consequence, BI gravity seems to forbid the possibility of packing
energy in arbitrarily small regions. Differing from GR, any
junction of the vacuum solution (\ref{metcuerda}) with an inner
solution has to be made at a radius bigger than $\rho_o=J/(2M^2)$.

Finally, we can mention that the increasing of the deficit angle
(coming from the change $d\xi=Y(\rho)\, d\rho$ in
(\ref{metcuerda})) as the string is closer, might have important
observational implications on the lensing effect.

Additional solutions for a wider set of parameters $(A,B)$, and
the search for non singular black hole fields coming from
(\ref{acciondet}), will be matter of future works.

\acknowledgements F.F. is indebted to Alina Fiorini for the
encouragement afforded during the time of writing these lines. We
would like to thanks G. Giribet for his valuable comments. This
research was supported by CONICET and Universidad de Buenos Aires.


\begin{thebibliography}{99}
\bibitem{Carlip} S. Carlip, \emph{Quantum Gravity in $2+1$ Dimensions},
Cambridge University Press (1998).

\bibitem{Brown} J.D. Brown, \emph{Lower Dimensional Gravity}, World
Scientific (1988).

\bibitem{Staru} A. Staruszkiewicz, Acta Phys. Polon. \textbf{24} (1963)
734.

\bibitem{Dilaton} R.V. Wagoner, Phys. Rev. \textbf{D1} (1970) 3209.

\bibitem{Dilaton1} S. Fernando, Phys. Lett. \textbf{B468} (1999) 201.

\bibitem{Dilaton2} T. Koikawa, T. Maki and A. Nakamula, Phys. Lett. \textbf{B414} (1997) 45.

\bibitem{Conformal1} G. Guralnik, A. Iorio, R. Jackiw and S.Y. Pi, Ann. Phys.
\textbf{308} (2003) 222.

\bibitem{Conformal2} D. Grumiller and W. Kummer, Ann. Phys. \textbf{308} (2003), 211.

\bibitem{Town1} E.A Bergshoef, O. Hohm and P. K. Townsend, Phys. Rev. Lett.
\textbf{102} (2009) 201301.

\bibitem{Town2} E.A Bergshoef, O. Hohm and P. K. Townsend, Phys. Rev. \textbf{D79} (2009)
124042.


\bibitem{TMG1} S. Deser, R. Jackiw and S. Templeton, Phys. Rev. Lett. \textbf{%
48} (1982) 975.

\bibitem{TMG2} S. Deser, R. Jackiw and S. Templeton, Ann. Phys. \textbf{140} (1982) 372.


\bibitem{TMG4} D.D.K. Chow, C.N. Pope and E. Sezgin, Class. Quant. Grav. \textbf{27} (2010) 105001.

\bibitem{H-E} S.W. Hawking and G.F. Ellis, \emph{The Large Scale
Structure of Space-Time}, Cambridge Univ. Press, (1973).

\bibitem{Vil} A. Vilenkin, Phys. Rept.
\textbf{121} (1985) 263.

\bibitem{Kib} M. B. Hindmarsh and T. W. B. Kibble, Rept. Prog. Phys. \textbf{58} (1995) 477.

\bibitem{Nos} R. Ferraro and F. Fiorini, Phys. Rev. \textbf{D75} (2007)
084031.

\bibitem{Nos2} R. Ferraro and F. Fiorini, Phys. Rev. \textbf{D78} (2008)
124019.

\bibitem{Born} M. Born, Nature \textbf{132} (1933) 282.

\bibitem{Borna} M. Born, Proc.
R. Soc. \textbf{A143} (1934) 410.

\bibitem{BornI} M. Born and L. Infeld, Proc. R. Soc. \textbf{A144} (1934)
425.

\bibitem{BornII} M. Born and L. Infeld,
\textbf{147} (1934) 522.

\bibitem{BornIII} M. Born and L. Infeld, \textbf{150} (1935) 141.

\bibitem{Tsey1} A.A. Tseytlin, \emph{Born-Infeld action, supersymmetry and string theory}, [arXiv:hep-th/9908105].

\bibitem{Tsey2} E. S. Fradkin and A. A. Tseytlin, Phys. Lett. \textbf{B163} (1985) 123.

\bibitem{deser3} S. Deser and G.W. Gibbons, Class. Quant. Grav. \textbf{15}
(1998) 35.

 \bibitem{Fein1} J. A. Feigenbaum, Phys. Rev. \textbf{D58} (1998)
124023.
\bibitem{Fein2} J. A. Feigenbaum, P.O. Freund and M. Pigli, Phys. Rev. \textbf{D57} (1998) 4738.

 \bibitem{Com1} D. Comelli, Phys. Rev. \textbf{D72}
(2005) 064018.

 \bibitem{Com2} D. Comelli and A. Dolgov, JHEP \textbf{0411} (2004)
062.

 \bibitem{Nieto} J. A. Nieto, Phys. Rev. \textbf{D70} (2004) 044042.

 \bibitem{Wohlfarth} M.N.R. Wohlfarth, Class. Quant. Grav. \textbf{21} (2004) 1927.

\bibitem{Quiros} R. Garc\'{\i}a-Salcedo, T. Gonzalez, C. Moreno, Y. Napoles, Y. Leyva and Israel Quiros, JCAP 1002:027 (2010).

 \bibitem{Tek1} I. Gullu, T. Cagri Sisman and B. Tekin, \emph{Born-Infeld extension of new massive gravity}, [arXiv:1003.3935].

 \bibitem{Tek2} I. Gullu, T. Cagri Sisman and B. Tekin, Phys. Rev. \textbf{D81} (2010) 104018.

\bibitem{Vollick} D. N. Vollick, Phys. Rev. \textbf{D72} (2005) 084026.

\bibitem{Max} M. Ba\~{n}ados, Phys. Rev. \textbf{D77} (2008) 123534.

\bibitem{Max1} M. Ba\~{n}ados and P. G. Ferreira \emph{Eddington's theory of gravity and its progeny}, [arXiv:1006.1769].

\bibitem{Hehl} J. Nitsch and F.W. Hehl, Phys. Lett. \textbf{B90} (1980) 98.
\bibitem{Hehl2} F.W. Hehl, J. D. McCrea, E. W. Mielke and Y. Ne'eman, Phy. Rept. \textbf{%
258} (1995) 1.

\bibitem{Maluf} J.W. Maluf, J. Math. Phys. \textbf{35} (1994) 335.

\bibitem{Nos3} F. Fiorini and R. Ferraro, Int. J. Mod. Phys. \textbf{A24}
(2009) 1686.

\bibitem{btz} M. Ba\~{n}ados, C. Teitelboim and J. Zanelli, Phys. Rev. Lett.
\textbf{69} (1992) 1849.

\bibitem{3D} S. Deser, R. Jackiw and G. 't Hooft, Ann. Phys. \textbf{152}
(1984) 220.

\bibitem{3D1} J. R. Gott, Phys. Rev. Lett. \textbf{66} (1991)
1126.

\bibitem{3D2} J.R. Gott and M. Alpert, Gen. Rel. Grav. \textbf{16}
(1984) 243.

\bibitem{3D3} S. Giddings, J. Abbot and K. Kuchar, Gen. Rel. Grav.
\textbf{16} (1984) 751.

\bibitem{3D4} S. Deser, R. Jackiw and G. 't Hooft, Phys. Rev. Lett. \textbf{68}
(1992) 267.

\end{thebibliography}
\end{document}